\pdfoutput=1
\documentclass[sigconf]{acmart}
\newtheorem*{key question}{\bf Key Question}

\newtheorem{corollary}{\bf Corollary}
\newtheorem{example}{\bf Example}
\newtheorem{definition}{\bf Definition}

\newtheorem{proposition}{\bf Proposition}
\newtheorem*{mechanism}{\bf Mechanism}
\newtheorem{thm}{\bf Theorem}
\newtheorem{lem}{\bf Lemma}
\usepackage{cases}
\usepackage{subfigure}
\newtheorem{problem}{\bf Problem}
\usepackage{dsfont}
\usepackage{multirow}
\usepackage{array} 
\usepackage{modroman}
\DeclareMathOperator*{\argmax}{argmax}
\usepackage{graphicx}
\usepackage{epstopdf}
\usepackage{tcolorbox}

\acmYear{2024}\copyrightyear{2024}
\setcopyright{acmlicensed}
\acmConference[MobiHoc '24]{International Symposium on Theory, Algorithmic Foundations, and Protocol Design for Mobile Networks and Mobile Computing}{October 14--17, 2024}{Athens, Greece}
\acmBooktitle{International Symposium on Theory, Algorithmic Foundations, and Protocol Design for Mobile Networks and Mobile Computing (MobiHoc '24), October 14--17, 2024, Athens, Greece}
\acmDOI{10.1145/3641512.3686394}
\acmISBN{979-8-4007-0521-2/24/10}

\acmConference[Mobihoc'24]{Mobihoc'24: ACM International Symposium on Theory, Algorithmic Foundations, and Protocol Design for Mobile Networks and Mobile Computing}{Oct. 14--Oct. 17, 2024}{Athens, Greece}
\begin{document}

\title{Social Welfare Maximization for Federated Learning with Network Effects}

\author{Xiang Li, Yuan Luo, Bing Luo, Jianwei Huang\let\thefootnote\relax\footnotetext{Xiang Li is with Shenzhen Institute of Artificial Intelligence and Robotics for Society (AIRS), and the School of Science and Engineering (SSE), The Chinese University of Hong Kong, Shenzhen (E-mail: xiangli2@link.cuhk.edu.cn). Yuan Luo is with SSE, AIRS, and Shenzhen Key Laboratory of Crowd Intelligence Empowered Low-Carbon Energy Network, The Chinese University of Hong Kong, Shenzhen (E-mail: luoyuan@cuhk.edu.cn). Bing Luo is with Data Science Research Center, Duke Kunshan University, Jiangsu (E-mail: bl291@duke.edu). Jianwei Huang is with SSE, AIRS, Shenzhen Key Laboratory of Crowd Intelligence Empowered Low-Carbon Energy Network, and CSIJRI Joint Research Centre on Smart Energy Storage, The Chinese University of Hong Kong, Shenzhen, Guangdong, 518172, P.R. China (Corresponding Author, E-mail: jianweihuang@cuhk.edu.cn).}}
\renewcommand{\shortauthors}{Xiang Li, et al.}

\begin{abstract}
  A proper mechanism design can help federated learning (FL) to achieve good social welfare by coordinating self-interested clients through the learning process. However, existing mechanisms neglect the network effects of client participation, leading to suboptimal incentives and social welfare. This paper addresses this gap by exploring network effects in FL incentive mechanism design. We establish a theoretical model to analyze FL model performance and quantify the impact of network effects on heterogeneous client participation. Our analysis reveals the non-monotonic nature of FL network effects. To leverage such effects, we propose a model trading and sharing (MTS) framework that allows clients to obtain FL models through participation or purchase. To tackle heterogeneous clients' strategic behaviors, we further design a socially efficient model trading and sharing (SEMTS) mechanism. Our mechanism achieves social welfare maximization solely through customer payments, without additional incentive costs. Experimental results on an FL hardware prototype demonstrate up to $148.86\%$ improvement in social welfare compared to existing mechanisms. 
\end{abstract}







\maketitle

\section{Introduction}
\begin{figure}[ht]
    \centering
    \includegraphics[width=0.4\textwidth]{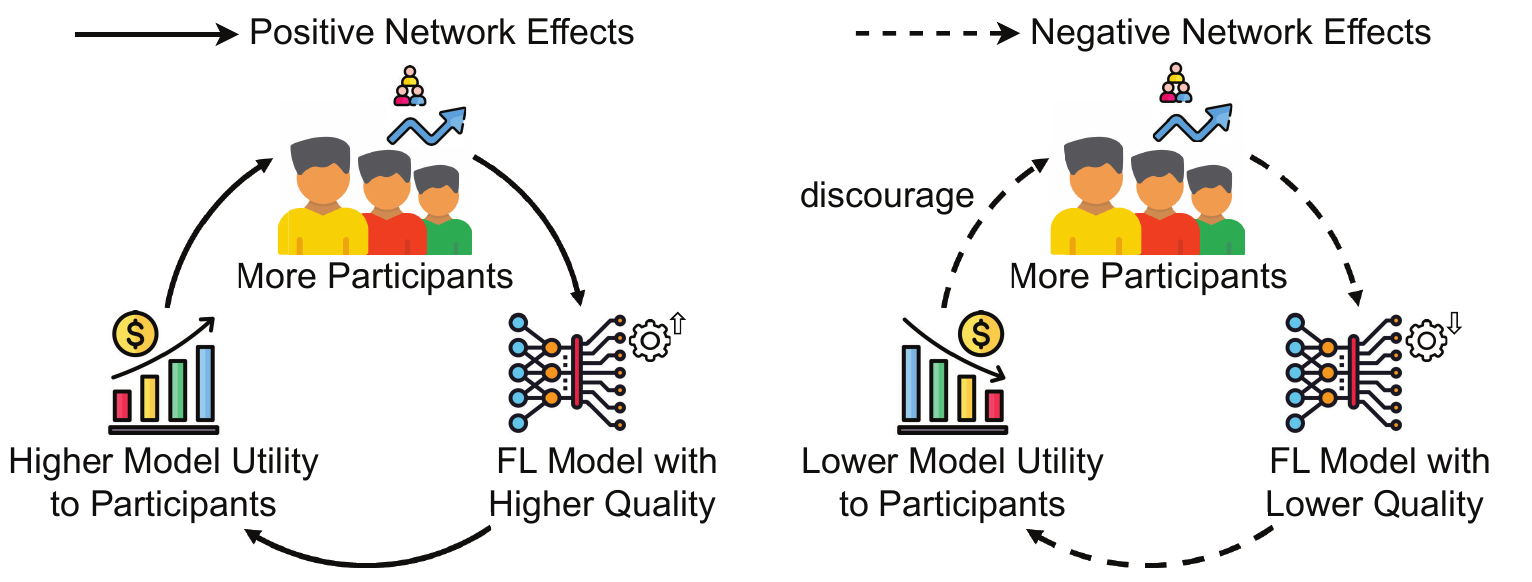}
    \vspace{-12pt}
    \caption{Network Effects of Client Participation in FL.}
    \label{NE}
    \Description{Network Effects of Client Participation in FL}
     \vspace{-10pt}
\end{figure}
Federated learning (FL) \cite{Zhan22,Tu22,Khan20} is an emerging machine learning (ML) paradigm that exploits massive data \cite{Nguyen2021-zp} and resources \cite{Guo2021-bf} across devices. As a collaborative approach, FL enables clients to enjoy the utility of employing well-trained models through participation, promising to deliver significant social welfare (i.e., \emph{the total payoffs for all parties involved} \cite{Rong21}). However, given the costs associated with model training, e.g., data labeling and computational costs \cite{zhao2023truthful}, self-interested clients may hesitate to participate and contribute their resources. Specifically, under existing FL frameworks (e.g., \cite{mcmahan2017communication,Tan2022,Abdul2021}), although clients can obtain and utilize the trained model through participation, the utility derived via employing the model could be insufficient to offset their costs, thereby discouraging their participation. With limited client participation, FL struggles to realize the full benefits of collaboration and maximize social welfare. Therefore, from the perspective of system efficiency \cite{Chinchuluun2008-pc}, designing a socially efficient (i.e., social welfare maximizing) mechanism is essential for FL to succeed.

Recent studies have made impressive progress in optimizing social welfare through mechanism design for FL, e.g., \cite{Thi21,Lin22,Lee20,Jiao21,Saputra23,Zhan20,Luo23,Zhan22}. However, they often neglect the \textbf{network effects} \cite{Easley2012-kj} arising from client participation. Specifically, \emph{network effects\footnote{In economics, network effect is a phenomenon where a product's utility varies with its client number, which often exhibits monotonicity, either positive (i.e., utility increases with clients, as in social networking services) or negative (e.g., network congestion).} refer to the interdependence between client participation, trained FL model performance, and corresponding utility, which vary with the network size (participant number)}. These effects influence the correlation between clients' participation decisions. For example, as more clients contribute to FL, training models with higher utility become more feasible, thus attracting even more participants in a positive cycle. Conversely, due to data heterogeneity \cite{luoyiqian}, more clients may also yield models with lower utility, discouraging client participation, as depicted in Figure \ref{NE}. Ignoring such network effects could lead to improper incentive mechanism design at different stages, either underpaying or overpaying clients, failing to adapt to the dynamics of clients' participation decisions. Motivated by this, we investigate the following key question in this paper:

\begin{key question}
\label{q1}
How to design a socially efficient mechanism for federated learning with network effects?
\end{key question}

To address the key question above, we propose a \underline{M}odel \underline{T}rading and \underline{S}haring (MTS) framework\begin{figure}[ht]
    \centering
    \includegraphics[width=0.41\textwidth]{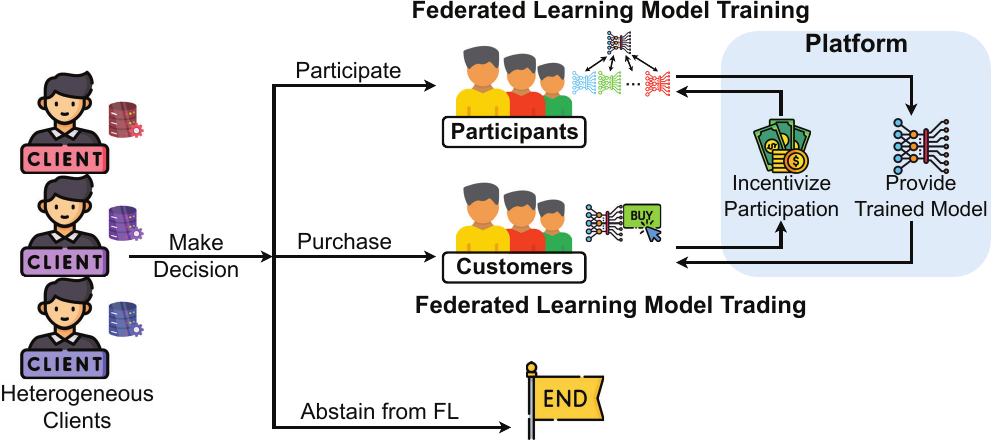}
    \vspace{-12pt}
    \caption{The Model Trading and Sharing (MTS) Framework.}
    \Description{The Model Trading and Sharing (MTS) Framework}
    \label{framework}
    \vspace{-10pt}
\end{figure}, inspired by the emerging ML model markets (e.g., \cite{StreamML,GravityAI}), to exploit network effects in FL and enhance social welfare. The MTS framework, as illustrated in Figure \ref{framework}, allows clients to obtain and utilize FL models in two ways: by participating in model training as a participant, or by purchasing trained models as a model customer. With this additional purchasing option, FL models are more accessible to clients (e.g., for those with high participation costs or limited data size), and payments from model customers can \emph{trigger network effects that promote client participation and increase social welfare}.

However, clients' strategic decisions prioritize optimizing their individual payoff and potentially deviate from socially optimal decisions. As a result, we need to further design a socially efficient mechanism for the MTS framework, while the unique characteristics of FL with network effects make it highly challenging. Specifically, due to the client data statistical heterogeneity (i.e., data size and distribution heterogeneity \cite{Yan22}) and the distributed nature of FL, quantifying the trained model performance requires complex theoretical models \cite{pmlr-v119-karimireddy20a}. This inherent complexity makes it difficult to evaluate network effects by determining the expected model utility and payoff that clients derive from participation. Furthermore, the privacy-preserving feature of FL restricts clients from accessing each other's private information, which leads to an incomplete information game, and the extra purchasing option expands clients' decision space even further.

Against this background, we first delve into the linear FL models, which are insightful for nonlinear models \cite{NIPS2017_6211080f,liang2020think,Trevor22,Mei22,AKrogh_1992}. Through the linear model setting and the quantification of network effects on FL model performance, we explore the Nash equilibrium \cite{mas1995microeconomic} among multiple clients with incomplete information. Building upon these, we design a \underline{S}ocially \underline{E}fficient \underline{M}odel \underline{T}rading and \underline{S}haring (SEMTS) mechanism that exploits the network effects of client participation. The SEMTS mechanism incentivizes heterogeneous clients to \emph{maximize social welfare using only payments from model customers, without incurring extra incentive costs}.

This paper advances the state of the art in the following ways:
\begin{itemize}
    \item \emph{The MTS Framework for FL:} To the best of our knowledge, this is the first study on social welfare optimization in FL with network effects. We propose the \underline{M}odel \underline{T}rading and \underline{S}haring (MTS) framework for FL to make models more accessible to clients and enhance social welfare. The MTS framework allows clients to obtain and utilize the trained global FL model either by participating in training or purchasing directly.
    \item \emph{Network Effects Analysis:} Our analysis reveals insights into network effects on heterogeneous client participation and model performance. Counter-intuitively, we show that FL network effects are not monotonic but potentially transit between positive and negative. In scenarios with low data distribution heterogeneity (i.e., non-i.i.d. level of client data), as more clients participate, FL model performance and utility decrease first and then increase. In contrast, with high data distribution heterogeneity, network effects of client participation will exhibit a reverse trend, initially positive but eventually turning negative.
    \item \emph{Socially Efficient Mechanism Design:} Based on the quantification of FL network effects, we derive clients' strategic decisions under incomplete information and incorporate them into the mechanism design for the MTS framework. Specifically, we design a \underline{S}ocially \underline{E}fficient \underline{M}odel \underline{T}rading and \underline{S}haring (SEMTS) mechanism. By exploiting the network effects of client participation, our designed mechanism maximizes social welfare, relying on payments received from model customers rather than incurring extra incentive costs. 
    \item \emph{Experimental Evaluation with Hardware Prototype:} We empirically evaluate our proposed mechanism and MTS framework on a hardware prototype of FL. The extensive experiments validate our analysis of network effects in FL and demonstrate that our mechanism can improve social welfare by up to $148.86\%$, compared to the theoretical upper bound of existing FL mechanisms that do not consider network effects.
\end{itemize}

The remainder of this paper is organized as follows. Section \ref{S1} reviews related works of mechanism design and network effects in FL. Section \ref{S2} introduces the system model and problem formulation. Section \ref{S3} presents our proposed MTS framework, which includes the mechanism design through the analysis of network effects. Section \ref{S4} provides the experimental evaluation on an FL hardware prototype. We conclude this paper in Section \ref{S5}. Due to the page limit, we leave detailed proof of our results and extra empirical experiments in the online appendix \cite{Online_appendix}.

\section{Related Works}\label{S1}
The mechanism design for social welfare optimization of FL (e.g., \cite{Thi21,Lin22,Lee20,Jiao21,Saputra23,Zhan20,Luo23,10476711,luoyiqian}) has made considerable advancements, as summarized in recent surveys \cite{Zhan22,Tu22,Khan20}. Specifically, Lee \emph{et al.} in \cite{Lee20} formulated the interaction of participants in FL as a Stackelberg game to maximize social welfare. Jiao \emph{et al.} in \cite{Jiao21} proposed an auction-based market model for incentivizing clients to participate in FL, which can efficiently maximize the social welfare. Saputra \emph{et al.} in \cite{Saputra23} developed a contract-based policy to increase the social welfare of vanilla FL methods. Zhan \emph{et al.} in \cite{Zhan20} evaluated ambiguous contributions in FL through a learning-based algorithm, which improves the overall payoff for participants. 

However, these works typically overlook the network effects in FL, which in practice could cause existing mechanisms to either underpay or overpay clients for incentivizing their participation. In contrast, this paper theoretically analyzes the network effects of client participation, which is relatively under-explored and more complex given the unique attributes of FL. By leveraging network effects, our designed mechanism can incentivize clients' participation to maximize the social welfare without requiring extra incentives.

The only paper that studies network effects in FL is \cite{hu2023} by Hu \emph{et al.}, which investigated the client’s evolving expectation of FL performance by modeling client behavior as a network effects game. However, this literature focused on the participation equilibrium analysis in FL, without fully considering client heterogeneity and social welfare. Specifically, they assumed that all clients share an expectation about the FL coalition and that client participation will always improve the trained model. In reality, clients often lack information about others' decisions, and FL model performance varies with the data statistical heterogeneity. This incomplete information and the impact of heterogeneity could lead their mechanism to fail to incentivize diverse clients for the desired equilibrium. In contrast, this paper designs a mechanism to incentivize the participation of heterogeneous clients with incomplete information, aiming to maximize the social welfare in FL.

\section{System Model and Problem Formulation}
\label{S2}
In this section, we begin by introducing the fundamentals of FL and modeling heterogeneous clients. Then, we specify our proposed MTS framework and formulate the mechanism design problem for social welfare optimization within it.

\subsection{Preliminaries of Federated Learning}
FL is a distributed ML paradigm that leverages multiple clients' resources to collaboratively train a shared model. To illustrate how FL works, we first provide an overview of the FL training process.

We consider a set of clients $\mathcal{N}=\{1, 2, \cdots, N\}$ interested in FL model training, which involves $N$ clients with heterogeneous characteristics. By participating in the training process, clients can obtain and utilize the trained global FL model. However, the FL model training process also incurs related costs for participants, e.g., due to resource consumption and data processing. Consequently, the self-interested $N$ clients may not fully participate in FL.

Let $\mathcal{K}\subseteq\mathcal{N}$ denote the set of participants involved in the FL training process. Each participant $k\in\mathcal{K}$ will utilize the private data set $\mathcal{D}_k$ of size $D_k$ to train a model. The local training objective of participant $k$ is to find the proper model parameters $\boldsymbol{w}$ that minimize the average loss $F_k(\boldsymbol{w})$ across all data samples $d\in\mathcal{D}_k$:
\begin{equation}
    F_k(\boldsymbol{w})=\frac{1}{D_k}\sum\nolimits_{d\in\mathcal{D}_k} f(\boldsymbol{w};d),
\end{equation}
where $f(\cdot)$ represents the per-sample loss function.

With locally trained models, FL aims to tackle the global loss optimization problem based on all participants' data, i.e., learn the optimal model parameters $\boldsymbol{w}^*$ to minimize the weighted average of all participants' local loss functions:
\begin{equation}
    \boldsymbol{w}^*=\arg\min_{\boldsymbol{w}} F(\boldsymbol{w})=\arg\min_{\boldsymbol{w}}\sum\nolimits_{k\in\mathcal{K}}\frac{D_k}{\sum_{k\in\mathcal{K}}D_k}F_k(\boldsymbol{w}).
\end{equation}

After introducing the preliminaries of FL, we then present heterogeneous clients interested in implementing FL.
\subsection{Heterogeneous Clients Modeling} 
This subsection describes heterogeneous characteristics of clients, along with their participation cost and model utility.

\subsubsection{Heterogeneous Clients}\label{Theoretical para} We study a set $\mathcal{N}=\{1,2,\cdots,N\}$ of $N$ clients with data statistical heterogeneity in FL, including both data distribution heterogeneity (i.e., non-i.i.d nature of client data) and data size heterogeneity.

\begin{itemize}
    \item To capture data distribution heterogeneity, as explored in \cite{Trevor22,NIPS2017_6211080f,liang2020think}, we consider the local feature represented by each client's data set independently follow a Gaussian distribution with diagonal covariances of $\sigma^2\in\mathbb{R}^d$, where $d$ is the feature dimension. With larger $\sigma^2$, the local features differ more across clients' respective data, indicating a higher client variance. Similarly, for the naturally occurring data variance \cite{montgomery2021introduction}, we incorporate it as noise $\delta\sim\mathcal{N}(0,\gamma^2)$ on the training targets, following a Gaussian distribution with variance $\gamma^2$. 
    \item Regarding data size heterogeneity of clients, we distinguish $N$ clients into a set $\mathcal{I}=\{1,2,\cdots,I\}$ of $I$ types by their different data sizes. Specifically, we refer to a client with data size $D_i$ as a type $i\in\mathcal{I}$ client. Let $\mathcal{N}_i$ denote the set of $N_i$ type-$i$ clients, where $\cup_{i\in\mathcal{I}}\mathcal{N}_i=\mathcal{N}$ and $\sum_{i\in\mathcal{I}}N_i=N$. Without loss of generality, we assume a larger index indicates a larger data size, i.e., $D_i\leq D_j$, if $i\leq j$, for any $i,j\in\mathcal{I}$.
\end{itemize}

\subsubsection{Participation Cost} For heterogeneous clients, participating in the FL model training process with their private data will incur a cost of resource consumption (e.g., computation and communication resources) and data processing (e.g., data labeling). This participation cost often positively\footnote{This positive correlation only aims to align with real-world situations, which will not affect our theoretical analysis and mechanism design later.} correlates with the client's data size, as illustrated in \cite{zhao2023truthful,karimireddy2022mechanisms}. Hence, we denote $C_i$ as the participation cost of type-$i$ clients, with $C_i\leq C_j$, if $i\leq j$, for any $i,j\in\mathcal{I}$.
\subsubsection{FL Model Utility} According to research on ML model markets \cite{Chen2019-js,10.14778/3447689.3447700,10.1145/3328526.3329589}, the client's utility of employing the FL model relies heavily on how well it performs on future (unseen) data, i.e., the model's generalization ability, quantified by the generalization error $\epsilon$. We thus define $U(\epsilon)\geq 0$ as the client's utility function of the FL model, which is non-increasing in $\epsilon$, implying that a lower generalization error corresponds to a higher model utility for clients. Additionally, from the practical correlation\footnote{Due to page limit, we focus on such typical utility functions. The analysis is the same for other utility functions, detailed in our appendix \cite{Online_appendix}.} between FL model performance and utility, e.g., \cite{Fernandez2020-cx,10.1145/3299869.3300078,Cong2022-pp}, as the model performance improves, generalization error reduction tends to be increasingly valuable to clients.
This trend continues until the generalization error $\epsilon$ reaches a certain threshold\footnote{As model improves, reducing $\epsilon$ is increasingly challenging and resource-intensive \cite{hestness2017deep}, making error reductions more valuable (higher marginal utility) for clients. Nevertheless, once the model is sufficient for its intended application, further improvements will yield diminishingly perceptible benefits.}, which we associate with feature variance $\sigma^2$, i.e., $(\epsilon-\sigma^2)U''(\epsilon)+2U'(\epsilon)\geq 0$, for all $\epsilon\neq\sigma^2$. 

Furthermore, as common in the literature \cite{Trevor22,liang2020think,Mei22}, we adopt a student-teacher setup for FL to theoretically quantify the generalization error $\epsilon$ of the trained global model. Specifically, given the feature dimension $d$, data variance $\gamma^2$, and client variance $\sigma^2$ discussed in Section \ref{Theoretical para}, we extend the analytical results in \cite{liang2020think} (which studied homogeneous FL scenarios, assuming uniform participant data sizes) to account for heterogeneous participants and derive the generalization error $\epsilon$ as follows:
\begin{equation}
    \label{E7}
        \epsilon=\frac{d\gamma^2}{K^2}\sum\nolimits_{i\in\mathcal{I}}\frac{K_i}{D_i}+\frac{K-1}{K}\sigma^2,
\end{equation}
where $K_i$ is the number of type-$i$ participants, for any $i\in\mathcal{I}$. We give the derivation of \eqref{E7} in the online appendix \cite{Online_appendix}.

From \eqref{E7}, the generalization error $\epsilon$ comprises two distinct components: $\frac{d\gamma^2}{K^2}\sum_{i\in\mathcal{I}}\frac{K_i}{D_i}$ and $\frac{K-1}{K}\sigma^2$, caused by data variance and client variance, respectively:

\begin{itemize}
    \item The generalization error arising from data variance $\gamma^2$ exhibits an inverse relationship with both the number $K$ of participants and the harmonic mean ${K}/({\sum_{i\in\mathcal{I}}K_i/D_i})$ of their individual data sizes.
    \item In contrast, the impact of client variance $\sigma^2$ on the generalization error $\epsilon$ increases with the number of participants. That is, although involving more participants with more local data reduces the generalization error $\epsilon$ attributed to data variance $\gamma^2$, it concurrently amplifies the error resulting from client variance $\sigma^2$.
\end{itemize}

In conclusion, according to $\epsilon$ in \eqref{E7}, the trained FL model performance depends on participants' number and data statistical heterogeneity, implying an interdependence between utility and clients' participation decisions, i.e., network effects. We also evaluate our theoretical results with nonlinear FL models through extensive experiments on an FL hardware prototype, elaborated in Section \ref{S4}.

\subsection{The MTS Frameworks of Federated Learning} 
While clients can enjoy FL model utility by participating in training, the associated participation costs may discourage some of them. To enhance model accessibility and exploit the network effects, we propose a Model Trading and Sharing (MTS) framework, which introduces an extra purchasing option for clients to obtain the trained model. In this subsection, we elaborate on the MTS framework, focusing on the decision space and the payoff of clients within it.

\subsubsection{Client Decision} Under the MTS framework, heterogeneous clients have three available strategies: abstain from obtaining the FL model, participate\footnote{To deliver clear insights, we focus on full participation behaviors of clients rather than partial ones (i.e., randomly joining some rounds of FL training), where the latter will yield similar analytical results, as studied in \cite{wang2024a,yang2021achievinglinearspeeduppartial}.} in the training process to obtain the FL model, or directly purchase the well-trained FL model. We denote the strategies of client $n\in\mathcal{N}_i$ with any type $i\in\mathcal{I}$ as $s_{n}\in\{A,J,B\}$, i.e.,
\begin{subnumcases}{\label{E3}s_{n}=}
    A, &\textrm{abstain from obtaining the FL model},\\
             J, &\textrm{join FL to obtain the model},\\
             B, &\textrm{buy the FL model}.
\end{subnumcases}

Given clients' decisions, we can calculate the total number of participants and FL model customers for any type $i\in\mathcal{I}$, defined as $K_i=\sum_{n\in\mathcal{N}_i}\mathds{1}_{s^*_{n}=J}$ and $B_i=\sum_{n\in\mathcal{N}_i}\mathds{1}_{s^*_{n}=B}$, respectively. The indicator function $\mathds{1}_{s^*_{n}=J}$ means that client $n$ decides to participate in FL model training, while $\mathds{1}_{s^*_{n}=B}$ indicates that client $n$ decides to purchase the trained FL model. Since the number of type-$i$ clients is $N_i$, we can also determine the number of clients with strategy $A$, which is $N_i-K_i-B_i$.

\subsubsection{FL Model Price and Participation Reward}\label{plat} As illustrated in Figure \ref{framework}, the MTS framework includes a platform that coordinates the FL model training and trading process. This platform enables heterogeneous clients to obtain the model by paying or, alternatively, by participating in training with a participation reward. We use $p$ to represent the price of the trained global FL model and denote $r_i$ as the participation reward for type-$i$ clients. In order to focus on the socially efficient mechanism design for FL with network effects, we assume such a platform is non-profit\footnote{For the platform operating on a cost-recovery basis, the analysis is the same but incorporates the platform's associated costs into social welfare.} and aims to maximize the social welfare. That is, the platform only leverages all payments received from model customers to incentivize client participation and model improvements in FL, i.e.,
\begin{equation}
    \sum\nolimits_{i\in\mathcal{I}}B_ip=\sum\nolimits_{i\in\mathcal{I}}K_ir_i.
\end{equation}

Note that the participation reward $r_{i}$, for any $i\in\mathcal{I}$, could be negative, which serves to penalize the participation of clients with poor contributions to the FL model performance and compensate other high-contributing participants.

\subsubsection{Client Payoff}
The payoff $\pi_{n}$ of a type-$i$ client $n\in\mathcal{N}_i$, for any $i\in\mathcal{I}$, within the MTS framework is a function of all clients' decisions, which also involves the model utility $U(\epsilon)$, participation cost $C_i$, participation reward $r_i$, and model price $p$. To simplify the notation, we denote the vector of all clients' strategies (i.e., $s_{n}$ for all $n\in\mathcal{N}$) as $\boldsymbol{s}$, and summarize the payoff function of client $n$ with type $i\in\mathcal{I}$ as follows:
\begin{subnumcases}{\label{E6}\pi_{n}(\boldsymbol{s})=}
    0,&\textrm{if $s_{n}=A$},\label{6a}\\
    U(\epsilon)-C_i+r_{i},&\textrm{if $s_{n}=J$},\label{6b}\\
    U(\epsilon)-p,&\textrm{if $s_{n}=B$}.\label{6c}
\end{subnumcases}

From (\ref{6a}), if client $n$ abstains from acquiring the trained global FL model, the payoff $\pi_n$ is zero. By participating in the FL model training process, the type-$i$ client $n\in\mathcal{N}_i$ will obtain the model utility $U(\epsilon)$ and participation reward $r_{i}$ while incurring the participation cost $C_i$, as stated in (\ref{6b}). When the client $n$ decides to purchase the trained FL model directly, the payoff $\pi_n$ is the difference between model utility $U(\epsilon)$ and model price $p$, as expressed in (\ref{6c}).

With the proposed MTS framework, we aim to optimize the social welfare in FL. Given the platform's non-profit nature, as detailed in Section \ref{plat}, the payoff of the platform will always be zero. That is, the objective of social welfare maximization is consistent with the maximization of clients' total payoff. To this end, we next formulate the mechanism design problem for social welfare optimization.

\subsection{Socially Efficient Mechanism Design Problem}
To maximize the FL social welfare, the mechanism design problem under the MTS framework entails optimizing both FL model price $p$ and participation reward $r_i$, for any $i\in\mathcal{I}$, as elaborated below.

For a set of clients $\mathcal{N}$ interested in the FL model under the MTS framework, the payoff of the platform is zero, and thus the social welfare $W_{\textrm{MTS}}$ is the sum of each client's payoff $\pi_{n}$ in \eqref{E6}, related to all clients' strategies. Given the number of participants $K_i$ and model customers $B_i$ of each client type $i\in\mathcal{I}$, we can express the social welfare $W_{\textrm{MTS}}$ as follows:
\begin{equation}\label{E5}
    W_{\textrm{MTS}}=\sum\nolimits_{i\in\mathcal{I}}[K_i(U(\epsilon)-C_i+r_{i})+B_i(U(\epsilon)-p)].
\end{equation}

According to $W_{\textrm{MTS}}$ in \eqref{E5} and considering clients' self-interested decision-making, i.e., making decisions that maximize their individual payoff, we formulate the mechanism design problem of social welfare optimization as Problem \ref{Pb1}.
\begin{tcolorbox}
\begin{problem}[Socially Efficient Mechanism Design]\label{Pb1}
    \begin{align*}
    \max_{p,r_i}\textrm{ }&\sum\nolimits_{i\in\mathcal{I}} [K_i(U(\epsilon)-C_i+r_{i})+B_i(U(\epsilon)-p)],\\
    s.t.\textrm{ }&\epsilon=\frac{d\gamma^2}{K^2}\sum\nolimits_{i\in\mathcal{I}}\frac{K_i}{D_i}+\frac{K-1}{K}\sigma^2,\\
    &\sum\nolimits_{i\in\mathcal{I}}B_ip=\sum\nolimits_{i\in\mathcal{I}}K_ir_i,\\
    &s^*_{n}=\argmax_{s_{n}\in\{A,J,B\}} \pi_{n}(\boldsymbol{s}),\textrm{ }\textrm{ }\forall n\in\mathcal{N}_i,\forall i\in\mathcal{I},\\
    &\pi_{n}(\boldsymbol{s}^*)\geq \pi_{n}(s_{n},\boldsymbol{s}^*\setminus\{s^*_{n}\}),\textrm{ }\textrm{ }\forall n\in\mathcal{N}_i,\forall i\in\mathcal{I},\\
    &K_i=\sum\nolimits_{n\in\mathcal{N}_i}\mathds{1}_{s^*_{n}=J},\textrm{ }\textrm{ }\forall i\in\mathcal{I},\\
    &B_i=\sum\nolimits_{n\in\mathcal{N}_i}\mathds{1}_{s^*_{n}=B},\textrm{ }\textrm{ }\forall i\in\mathcal{I}.
\end{align*}
\end{problem}
\end{tcolorbox}

In the following section, we will design a socially efficient mechanism under the MTS framework to solve Problem \ref{Pb1}. The mechanism takes into account both the FL model performance and the number of benefiting clients, leveraging the network effects of client participation to maximize the social welfare.
\section{Mechanism Design with Network Effects}
\label{S3}
We first theoretically analyze the FL network effects. Then, building upon this analysis, we study the social efficiency of the MTS framework and propose the mechanism that maximizes social welfare.
\subsection{Network Effects of Client Participation}\label{S4,1}
FL enables clients to collaboratively train and share a model with the aim of generating accurate outputs for a particular task. Based on the generalization error $\epsilon$ derived in \eqref{E7}, we begin by exploring the impact of client collaboration in FL, as summarized in Corollary \ref{C1}, in order to understand the network effects of client participation.
\begin{corollary}
\label{C1}
    Consider two disjoint non-empty participant coalitions $\mathcal{K}_a$ and $\mathcal{K}_b$, where $H_a$ and $H_b$ are the harmonic means of participants' data sizes in $\mathcal{K}_a$ and $\mathcal{K}_b$, with $H_a\leq H_b$. The generalization error $\epsilon_a$ and $\epsilon_b$ of the model trained by $\mathcal{K}_a$ and $\mathcal{K}_b$ satisfies $\epsilon_{a+b}<\max\{\epsilon_a,\epsilon_b\}$, if and only if $\frac{\sigma^2}{\gamma^2}<\frac{H_bK_b+K_a(2H_b-H_a)}{H_aH_b(K_a+K_b)/d}$, where $\epsilon_{a+b}$ is the generalization error of the model trained by $\mathcal{K}_a\cup\mathcal{K}_b$.
\end{corollary}

Corollary \ref{C1} implies that client collaboration in FL can benefit individual participants (or coalitions of multiple participants) by enabling them to obtain a model with lower generalization error $\epsilon$ compared to training alone. Nevertheless, collaboration is not always beneficial to all participants, especially when client variance $\sigma^2$ outweighs data variance $\gamma^2$ beyond a certain threshold. In such cases, the participation of a client could increase the generalization error of the FL model trained by existing participants. This decline in model performance reduces model utility for existing participants, exhibiting negative network effects of client participation in FL.

To thoroughly investigate the varying network effects of client participation, we quantify it in terms of the change in the generalization error $\epsilon$ of the trained FL model caused by client participation, as defined in Definition \ref{NECP}. Then, we derive conditions under which FL network effects are non-negative, established in Theorem \ref{T1}.

\begin{definition}[Network Effects \cite{Easley2012-kj} of Client Participation]
    \label{NECP} Given any non-empty coalition of existing participants $\mathcal{K}$, the participation of a client $n$ results in network effects manifested as the difference between the generalization error of the FL model trained by $\mathcal{K}$ and the generalization error of the model trained by $\mathcal{K}\cup\{n\}$, expressed as $\epsilon(\mathcal{K})-\epsilon(\mathcal{K}\cup\{n\})$.
\end{definition}

\begin{thm}
    \label{T1}
    Consider any non-empty coalition of existing participants $\mathcal{K}$, the participation of a client $n$ with data size $D$ will bring non-negative network effects, i.e., $\epsilon(\mathcal{K})-\epsilon(\mathcal{K}\cup\{n\})\geq 0$, if and only if ${1}/{D}$ is no more than $\eta$ in \eqref{delta}, i.e., ${1}/{D}\leq \eta$,
    \begin{equation}\label{delta}
        \eta=\frac{(2K+1)\sum_{i\in\mathcal{I}}{K_i}/{D_i}}{K^2}-\frac{(K+1)\sigma^2}{d\gamma^2K},
    \end{equation}
where $K_i$ is the number of type-$i$ participants in $\mathcal{K}$, for any $i\in\mathcal{I}$, and $K=\sum_{i\in\mathcal{I}}K_i$.
\end{thm}

We give the detailed proof of Theorem \ref{T1} in the online appendix \cite{Online_appendix}.
With $\eta$ in \eqref{delta}, Theorem \ref{T1} identifies two factors that determine the network effects of client participation: the data size of the new participant against that of existing participants, and the ratio of client heterogeneity $\sigma^2$ to data heterogeneity $\gamma^2$. 
\begin{itemize}
    \item If the ratio of client heterogeneity to data heterogeneity ${\sigma^2}/{\gamma^2}$ reaches the threshold $\frac{(2K+1)\sum_{i\in\mathcal{I}}{K_i}/{D_i}}{K(K+1)/d}$, client participation will always increase the generalization error $\epsilon$ of the trained FL model, resulting in negative network effects. 
    \item Conversely, when ${\sigma^2}/{\gamma^2}$ is less than $\frac{(2K+1)\sum_{i\in\mathcal{I}}{K_i}/{D_i}}{K(K+1)/d}$, a new participant needs only one-third of the harmonic mean of existing participants' respective data sizes to positively impact the current model (i.e., reducing generalization error $\epsilon$), as deduced by the upper bound $\frac{3\sum_{i\in\mathcal{I}}K_iD_i}{K}$ on $\eta$ in \eqref{delta}. 
\end{itemize}

However, the composition of participants changes with the participation of new clients, which in turn will affect the specific data size threshold required for future new clients to ensure non-negative network effects from their participation.

Given the above discussions, heterogeneous participants are intertwined to affect the performance of the FL model and the network effects of client participation. To demonstrate how FL model performance and network effects vary with the number of participants from different types, we first take $I=2$ types of clients, each possessing homogeneous data sets (i.e., $\sigma^2=0$), as an illustrative example, shown in Example \ref{E1}.
\begin{example}[Two Types of Clients]
\label{E1}
    Consider $I=2$ types of clients with $\sigma^2=0$ in FL, and a coalition of participant $\mathcal{K}$. As the number $K_1$ of type-$1$ participants in $\mathcal{K}$ increases, the generalization error $\epsilon$ of the trained global FL model initially increases, but eventually decreases once $K_1\geq\lceil{({-2K_2D_1-D_2+\sqrt{4K_2^2(D_2-D_1)^2+D_2^2}})/{2D_2}}\rceil$. In contrast, the generalization error consistently decreases as the number $K_2$ of type-$2$ participants in $\mathcal{K}$ grows.
\end{example}
Example \ref{E1} provides an interesting insight under the i.i.d. data distribution: when the number of participants with small data size falls below a certain threshold (as referred to type-$1$ clients in Example \ref{E1}), their participation will incur negative network effects; however, once the number of these participants exceeds the threshold, their participation will yield positive network effects. In contrast, the participation of clients with the largest data size brings positive network effects, regardless of the number of such participants. 

After introducing the case of two client types and i.i.d. data distribution, we extend our analysis to general scenarios with $I$ client types and non-i.i.d. data distribution, as elaborated in Theorem \ref{T2}.

\begin{thm}
\label{T2}
    Consider a coalition of existing participants $\mathcal{K}\neq\emptyset$, where $K_i$ is the number of type-$i$ participants in $\mathcal{K}$, for any $i\in\mathcal{I}$, and $K=\sum_{i\in\mathcal{I}}K_i$. Then, for clients with type $j\in\mathcal{I}$,
    \begin{itemize}
    \item if ${1}/{D_j}<{\sigma^2}/{d\gamma^2}$ and ${1}/{D_j}\leq\eta$, network effects of client participation are initially non-negative, but eventually become negative as more type-$j$ clients participate.
    \item if ${\sigma^2}/{d\gamma^2}\leq{1}/{D_j}\leq\eta$, network effects of client participation are always non-negative.
    \item if $\eta<{1}/{D_j}\leq{\sigma^2}/{d\gamma^2}$, network effects of client participation are always negative.
    \item if $\max\{{\sigma^2}/{d\gamma^2},\eta\}<{1}/{D_j}$, network effects of client participation are initially negative, but eventually become positive as more type-$j$ clients participate.
\end{itemize}
\end{thm}

\begin{figure}[ht]\vspace{-6pt}
    \centering
    \includegraphics[width=0.32\textwidth]{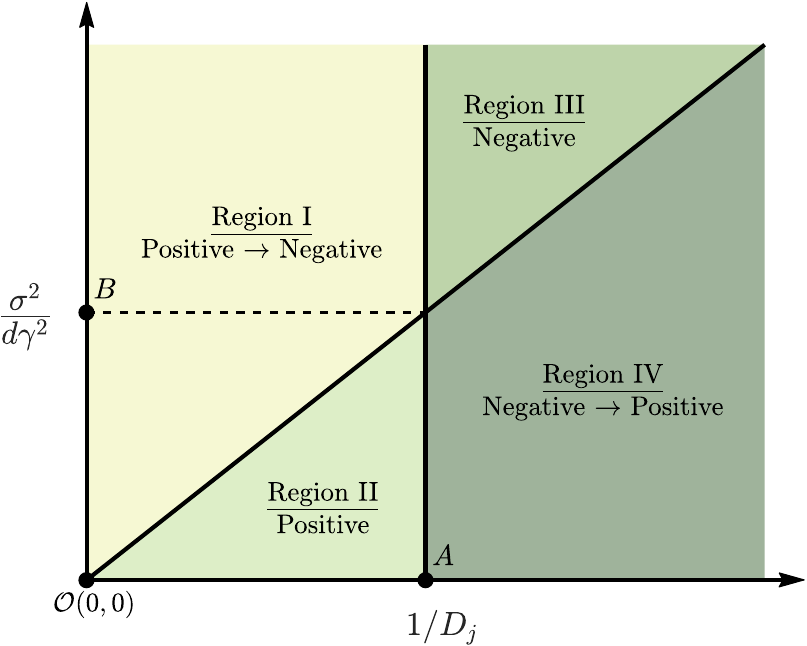}
    \boxed{
\begin{array}{ll}
\text{A}=({((2K+1)\sum_{i\in\mathcal{I}}{K_i}/{D_i}})/{K^2}-{(K+1)\sigma^2}/{d\gamma^2K},0).\\
\text{B}=(0,{((2K+1)\sum_{i\in\mathcal{I}}{K_i}/{D_i}})/{K^2}-{(K+1)\sigma^2}/{d\gamma^2K}).
\end{array}}
    \vspace{-8pt}
    \caption{Network Effects of Type-j Client Participation.}
    \Description{Network Effects of Type-j Client Participation}
    \label{Region}
    \vspace{-8pt}
\end{figure}

We give the proof of Theorem  \ref{T2} in online appendix \cite{Online_appendix}. From Theorem \ref{T2}, network effects of type-$j$ client participation not only depend on heterogeneity ratio $\sigma^2/\gamma^2$ and data size $D_j$, but also potentially vary with the number of type-$j$ participants, for any $j\in\mathcal{I}$. Specifically, we can divide the network effects of client participation into four regions, as depicted in Figure \ref{Region}. In Region \nbRoman{1}, the trained FL model performance (and model utility) first increases and then decreases with the growing number of participants of the same type. In contrast, for Region \nbRoman{2}, the network effects are always positive, reducing the generalization error $\epsilon$ of the FL model. Region \nbRoman{3} and Region \nbRoman{4}, however, exhibit reverse trends of network effects in Region \nbRoman{2} and Region \nbRoman{1}, respectively. As a result, given any participant coalition $\mathcal{K}\neq\emptyset$ and heterogeneity ratio $\sigma^2/\gamma^2$, we can determine the network effects for different client types. For example, when $\sigma^2=0$ (representing an i.i.d. data distribution, as indicated by line $OA$ in Figure \ref{Region}),  the network effects will correspond to Regions \nbRoman{3} or \nbRoman{4}, consistent with our observation in Example \ref{E1}.

So far, through the theoretical analysis of FL model performance, we have investigated network effects arising from the participation of heterogeneous clients. Next, we delve into the social efficiency of the MTS framework, in order to develop a corresponding mechanism that maximizes social welfare.

\subsection{Social Efficiency of the MTS Framework}
The social welfare \cite{Rong21} characterizes the total payoff for all parties involved. Under the MTS framework, the platform's payoff is always zero, detailed in Section \ref{plat}, and thus, the social welfare is the sum of each client's payoff. As formulated by $W_{\textrm{MTS}}$ in \eqref{E5}, the social welfare of the MTS framework depends on each client decision $s_n$ and the resultant FL model performance (quantified by generalization error). Through the generalization error $\epsilon$ of the FL model derived in \eqref{E7}, $W_{\textrm{MTS}}$ is a function of the number of heterogeneous participants and model customers, corresponding to different FL social states \cite{Easley2012-kj} introduced in Definition \ref{D1}.

\begin{definition}[Social State \cite{Easley2012-kj} of FL]
    \label{D1} Consider a group of clients with a set of pure strategies $\{A,J,B\}$ in \eqref{E3}. A social state is a vector $\{\boldsymbol{K},\boldsymbol{B}\}$ gives the number $K_i$ and $B_i$ of type-$i$ clients with strategy $J$ and $B$ respectively, for all $i\in\mathcal{I}$. 
\end{definition}

To determine optimal social welfare and design an appropriate mechanism, we explore the social efficiency of the MTS framework, with an emphasis on socially efficient states in Definition \ref{D2}. We begin by summarizing the features of socially efficient states under the MTS framework in Lemma \ref{L2}.

\begin{definition}[Socially Efficient States]\label{D2} Socially efficient states $\{\boldsymbol{K}^*,\boldsymbol{B}^*\}=\{K_i^*, B^*_i, \forall i \in\mathcal{I}\}$ are social states maximizing social welfare, i.e., $W_\textrm{MTS}(\boldsymbol{K}^*,\boldsymbol{B}^*)=\max W_\textrm{MTS}(\boldsymbol{K},\boldsymbol{B})$.
\end{definition}

\begin{lem}
\label{L2}
    In socially efficient states under the MTS framework, either no clients obtain the FL model, i.e., $\sum_{i\in\mathcal{I}}(K^*_i+B^*_i)=0$, or all of them obtain it, i.e., $\sum_{i\in\mathcal{I}}(K^*_i+B^*_i)=N$. 
\end{lem}

Lemma \ref{L2} implies the only two possibilities for the socially efficient states under the MTS framework. If the optimal social welfare $W_{\textrm{MTS}}^*$ is zero, the minimum possible value, then $\{\boldsymbol{K}^*,\boldsymbol{B}^*\}=\boldsymbol{0}$ must be a socially efficient state, that is, $\sum_{i\in\mathcal{I}}(K^*_i+B^*_i)=0$. When the optimal social welfare $W_{\textrm{MTS}}^*$ under the MTS framework is larger than zero, the socially efficient states will satisfy $\sum_{i\in\mathcal{I}}(K^*_i+B^*_i)=N$, indicating that all clients will obtain the trained global FL model, either by participating in training or purchasing it directly.

Building upon the above discussions, we proceed to study the socially efficient states with positive optimal social welfare, which is the transformation of Problem \ref{Pb1} and aligns with its objective function. We formulate such a socially efficient states problem as Problem \ref{Pb2}, incorporating the generalization error $\epsilon$ in \eqref{E7}. The socially efficient states addressed in Problem \ref{Pb2} will indicate strategies that mechanism for solving Problem \ref{Pb1} aims to incentivize clients to adopt. This establishes the foundation for us to design the mechanism that optimizes social welfare under the MTS framework later. 
\begin{tcolorbox}
\begin{problem}[Socially Efficient States of Problem \ref{Pb1}]\label{Pb2}
    \begin{align*}
    \max\quad &NU\left(\frac{d\gamma^2}{K^2}\sum\nolimits_{i\in\mathcal{I}}\frac{K_i}{D_i}+\frac{K-1}{K}\sigma^2\right)-\sum\nolimits_{i\in\mathcal{I}}K_iC_i,\\
    s.t.\quad&K=\sum\nolimits_{i\in\mathcal{I}}K_i,\\
    &0\leq K_i+ B_i\leq N_i,\quad\forall i\in\mathcal{I},\\
    var.\quad &K_i\in\mathbb{N},\textrm{ } B_i\in\mathbb{N},\quad\forall i\in\mathcal{I}.
\end{align*}
\end{problem}
\end{tcolorbox}

When dealing with Problem \ref{Pb2}, due to the varying network effects, we divide the type set $\mathcal{I}$ of heterogeneous clients into two distinct subsets: $\mathcal{I}_L$ (low type index) and $\mathcal{I}_H$ (high type index).  Such categorization is based on two key factors that affect the network effects of client participation, i.e., client data size $D_i$ and heterogeneity ratio $\sigma^2/\gamma^2$, as elaborated in Theorem \ref{T2}. For clients with type $i\in\mathcal{I}_L$, their data size satisfies that $D_i\leq d\gamma^2/\sigma^2$, whereas the data size of clients with type $j\in\mathcal{I}_H$ is larger than $d\gamma^2/\sigma^2$. By separately analyzing these two sets of clients, we determine the participation pattern of clients in the socially efficient states that yield positive optimal social welfare $W_{\textrm{MTS}}^*>0$, concluded in Proposition \ref{P1}. 

\begin{proposition}\label{P1}
Consider any non-empty set of client types $\mathcal{I}=\mathcal{I}_{L}\cup\mathcal{I}_{H}$, where $D_i\leq {d\gamma^2}/{\sigma^2}$ for each $i\in \mathcal{I}_{L}$, and $D_j> {d\gamma^2}/{\sigma^2}$ for each $j\in \mathcal{I}_{H}$. In socially efficient states with positive optimal social welfare $W_{\textrm{MTS}}^*>0$,

\begin{itemize}
    \item if $\mathcal{I}_H=\emptyset$, then $K_i^*\in\{0,N_i\}$, $\forall i\in\mathcal{I}_L$.
    \item if $\mathcal{I}_H\neq\emptyset$, then $K_i^*=0$, $\forall i\in\mathcal{I}_L$, and $K_j^*\in\{0,N_j\}\cup\{K_j\in\mathbb{N}:{\partial W_{\textrm{MTS}}(K_j)}/{\partial K_j}=0\}$, $\forall j\in\mathcal{I}_H$.
\end{itemize}
\end{proposition}

Proposition \ref{P1} reveals that socially efficient states typically exhibit an ``all-or-none'' participation pattern for any client type. However, for clients with type $j \in \mathcal{I}_H$, partial participation in training can also lead to socially efficient states. The underlying rationale is consistent with Theorem \ref{T2}. Specifically, according to Figure \ref{Region}, clients of type $i\in\mathcal{I}_L$ will fall into Region \nbRoman{2} and Region \nbRoman{4}, while clients of type  $j\in\mathcal{I}_H$ will occupy Region \nbRoman{1} and Region \nbRoman{3}. Consequently, in socially efficient states, type-$i$ clients will either all participate in training to exploit the positive network effects, or all abstain from training to avoid incurring negative network effects. In contrast, clients of type $j\in\mathcal{I}_H$ demonstrate an additional possibility for socially efficient states: only part of the clients participate in training, sufficient to reap positive network effects but not so much as to trigger negative ones (i.e., reaching the threshold of participant number $K^*_j=\{K_j\in\mathbb{N}:{\partial W_{\textrm{MTS}}(K_j)}/{\partial K_j}=0\}$ that network effects turn from positive to negative). 

To summarize, in scenarios with low data distribution heterogeneity (where $\sigma^2\leq d\gamma^2/D_I$, implying that $\mathcal{I}_H=\emptyset$), clients of the same type must make consistent participation decisions to maximize the social welfare under the MTS framework, i.e., $K_i^*=0$ or $K^*_i=N_i$, for any type $i\in\mathcal{I}_L$. Conversely, with high data distribution heterogeneity (where $\sigma^2>d\gamma^2/D_I$, implying that $\mathcal{I}_H\neq\emptyset$), socially efficient states will not involve participants of type $i\in\mathcal{I}_L$, and only clients of type $j\in\mathcal{I}_H$ potentially participate in FL.

After investigating socially efficient states $\{\boldsymbol{K}^*,\boldsymbol{B}^*\}$, we compare the resultant optimal social welfare $W^*_{\textrm{MTS}}$ under our proposed MTS framework to that of the existing FL framework $W^*_{\textrm{FL}}$. Unlike the MTS framework, the existing FL framework (e.g., \cite{mcmahan2017communication,Tan2022,Abdul2021}) only allows clients to obtain and utilize the FL model by participating in the training process, resulting in the social welfare $W_{\textrm{FL}}$ below:
\begin{equation}
    W_{\textrm{FL}}(\boldsymbol{K})=\sum\nolimits_{i\in\mathcal{I}}K_i\left(U(\epsilon)-C_i\right).
\end{equation}

In contrast, the MTS framework enables clients to access the model through an extra purchasing option, which leads to the social welfare $W_{\textrm{MTS}}$ in \eqref{E5}. As a result, our proposed MTS framework ensures that the optimal social welfare $W^*_{\textrm{MTS}}$ is no less than that of the existing FL framework $W^*_{\textrm{FL}}$, as presented in Proposition \ref{P2}.

\begin{proposition}\label{P2}
Given any client set $\mathcal{N}$, the optimal social welfare achieved by the MTS framework $W^*_{\textrm{MTS}}$ is no less than that of the existing FL framework $W^*_{\textrm{FL}}$, i.e., $W^*_{\textrm{MTS}}\geq W^*_{\textrm{FL}}$.
\end{proposition}

Following the analysis of social efficiency under our proposed MTS framework, we next address its practical implementation. We aim to design a mechanism for the MTS framework that leads self-interested clients, who prioritize optimizing their individual payoffs and thus potentially deviate from socially efficient states, to maximize the social welfare. 

\subsection{Socially Efficient Mechanism Design}
The problem of socially efficient mechanism design under the MTS framework, as formulated in Problem \ref{Pb1}, entails the optimization of FL model price $p$ and the participation reward $r_i$ for type-$i$ clients. 

From Lemma \ref{L2} and Proposition \ref{P1}, socially efficient states, due to network effects, require clients of each type $i\in\mathcal{I}$ to transit between specific social states in a way that maximizes social welfare. For instance, in scenarios with low data distribution heterogeneity $\sigma^2\leq d\gamma^2/D_I$, the socially efficient state is $K_i^*\in\{0,N_i\}$. However, each client $n\in\mathcal{N}_i$, irrespective of their type $i\in\mathcal{I}$, will seek to make decision $s^*_{n}$ that optimizes individual payoff $\pi^*_{n}$ in \eqref{E6} under the MTS framework, rather than social welfare. Meanwhile, the initial negative network effects, as elaborated in Theorem \ref{T2}, also prevent self-interested clients from reaching socially efficient states, thereby significantly complicating Problem \ref{Pb1}. In response to these challenges, we propose a \underline{S}ocially \underline{E}fficient \underline{M}odel \underline{T}rading and \underline{S}haring (SEMTS) mechanism, which aligns client payoff with social welfare while accounting for the network effects of client participation, as detailed below.

To maximize the social welfare under the MTS framework, we incentivize heterogeneous clients to overcome the negative network effects phase and exploit positive network effects. Specifically, corresponding to Proposition \ref{P1}, the SEMTS mechanism includes considerations based on distinct data distribution heterogeneity (i.e., different non-i.i.d. levels). For scenarios with low data distribution heterogeneity $\sigma^2\leq d\gamma^2/D_I$,  we construct a function $\theta(\boldsymbol{K})$ in \eqref{E8} for all clients in the intermediate social states (i.e., $0<K_i<N_i$, for any $i\in\mathcal{I}$), defined as follows:
\begin{align}\label{E8}
    \theta(\boldsymbol{K})=&\sum\nolimits_{i\in\mathcal{I}}\left(\right.\frac{N_i-K_i}{N_iI}U(\epsilon(K_i=0))\nonumber\\
    &-\frac{K_i}{N_iI}U(\epsilon(K_i=N_i))-\frac{K_iC_i}{N}\left.\right).
\end{align}

The function $\theta(\boldsymbol{K})$ in \eqref{E8} derives the average payoff of clients in socially efficient states and makes transitions among intermediate social states consistent with individual payoff maximization. In particular, when the participation of clients contributes to achieving socially efficient states, their payoff with the structure in \eqref{E8} will ensure positive network effects (i.e., increase the individual payoff of all participants), and vice versa. In contrast, for scenarios with high data distribution heterogeneity $\sigma^2> d\gamma^2/D_I$, the SEMTS mechanism focuses on ensuring a non-negative payoff for each participant, in order to initiate positive network effects. We conclude the details of the SEMTS mechanism in the following, which avoids discouraging client participation due to temporary negative network effects or negative payoff.

\begin{mechanism}[The SEMTS Mechanism]\label{M1}
Consider any set of client types $\mathcal{I}=\mathcal{I}_{L}\cup\mathcal{I}_{H}$, where $D_i\leq {d\gamma^2}/{\sigma^2}$ for each $i\in \mathcal{I}_{L}$, and $D_j> {d\gamma^2}/{\sigma^2}$ for each $j\in \mathcal{I}_{H}$. The SEMTS mechanism is as follows:

(1) For the scenario with low data distribution heterogeneity, i.e., $\sigma^2\leq d\gamma^2/D_I$, implying that $\mathcal{I}_H=\emptyset$:

    \begin{itemize}
    \item Model Price $p$:
    \end{itemize} 
    \begin{subnumcases}{\label{E9}p=}
        U(\epsilon), &$\textrm{if }\textrm{ }\theta(\boldsymbol{K})<0$,\\
        U(\epsilon)-\theta(\boldsymbol{K}), &$\textrm{if }\textrm{ }\theta(\boldsymbol{K})\geq0$.
    \end{subnumcases}
    \begin{itemize}
    \item Participation Reward $r_{i}$, for any $i\in\mathcal{I}$:
\end{itemize}
\begin{subnumcases}{\label{E10}r_{i}=}
        C_i-\frac{\sum\nolimits_{j\in\mathcal{I}}K_jC_j}{N}, &$\textrm{if }\textrm{ } \theta(\boldsymbol{K})<0, W^*_{\textrm{MTS}}\leq 0$,\\
        C_i-U(\epsilon), &$\textrm{if }\textrm{ } \theta(\boldsymbol{K})<0, W^*_{\textrm{MTS}}> 0$,\\
        C_i-U(\epsilon)+\theta(\boldsymbol{K}), &$\textrm{if }\textrm{ } \theta(\boldsymbol{K})\geq0$.
    \end{subnumcases}
    
(2) For the scenario with high data distribution heterogeneity, i.e., $\sigma^2>d\gamma^2/D_I$, implying that $\mathcal{I}_H\neq\emptyset$:
    \begin{itemize}
    \item Model Price $p$:
    \end{itemize}
    \begin{subnumcases}{\label{E11}p=}
        U(\epsilon), &$\textrm{if }\textrm{ }U(\epsilon)<\frac{\sum\nolimits_{j\in\mathcal{I}}K_jC_j}{N}$,\\
        \frac{\sum\nolimits_{j\in\mathcal{I}}K_jC_j}{N}, &$\textrm{if }\textrm{ }U(\epsilon)\geq\frac{\sum\nolimits_{j\in\mathcal{I}}K_jC_j}{N}$.
    \end{subnumcases}
    \begin{itemize}
    \item Participation Reward $r_{i}$, for any $i\in\mathcal{I}$:
\end{itemize}
\begin{subnumcases}{\label{E12}r_{i}=}
        C_i-U(\epsilon), &$\textrm{if }\textrm{ } U(\epsilon)<\frac{\sum\nolimits_{j\in\mathcal{I}}K_jC_j}{N}, W^*_{\textrm{MTS}}> 0$,\\
        C_i-\frac{\sum\nolimits_{j\in\mathcal{I}}K_jC_j}{N}, &$\textrm{otherwise}$.
    \end{subnumcases}

\end{mechanism}

With the model price $p$ and participation reward $r_i$ designed in the SEMTS mechanism, given any social state, the optimal decisions of all self-interested clients align with maximizing social welfare. Each client will make the decision to promote social welfare, i.e., as a participant or model customer that leads to socially efficient states, regardless of any other client's decision (after considering the coupling with other clients). Furthermore, once clients reach socially efficient states, they benefit from positive network effects without relying on any constructed function, e.g., $\theta(\boldsymbol{K})$. This implies that our mechanism can operate solely on payments from FL model customers rather than requiring extra incentives. We summarize the properties of the SEMTS mechanism in Theorem \ref{T3}, and give the detailed proof in the online appendix \cite{Online_appendix}.

\begin{thm}
    \label{T3}
    The SEMTS mechanism leads to the socially efficient states $\{\boldsymbol{K}^*,\boldsymbol{B}^*\}$ and achieves optimal social welfare $W_{\textrm{MTS}}^*$ under the MTS framework, which operates without incurring extra incentive costs, i.e., $\sum_{i\in\mathcal{I}}B_ip=\sum_{i\in\mathcal{I}}K_ir_i$.
\end{thm}

Overall, we have analyzed the network effects of client participation in FL and established an SEMTS mechanism for maximizing social welfare under the MTS framework. Next, we will empirically evaluate our theoretical results through extensive experiments on an FL hardware prototype.

\section{Experimental Evaluation}
\label{S4}
This section begins by introducing the experimental setup. Then, we present empirical results to demonstrate our investigation of FL network effects and evaluate the social efficiency of our proposed SEMTS mechanism and MTS framework.

\subsection{Experimental Setup}
\subsubsection{Hardware Prototype of FL} 
\begin{figure}[ht]\vspace{-6pt}
    \centering
    \includegraphics[width=0.26\textwidth]{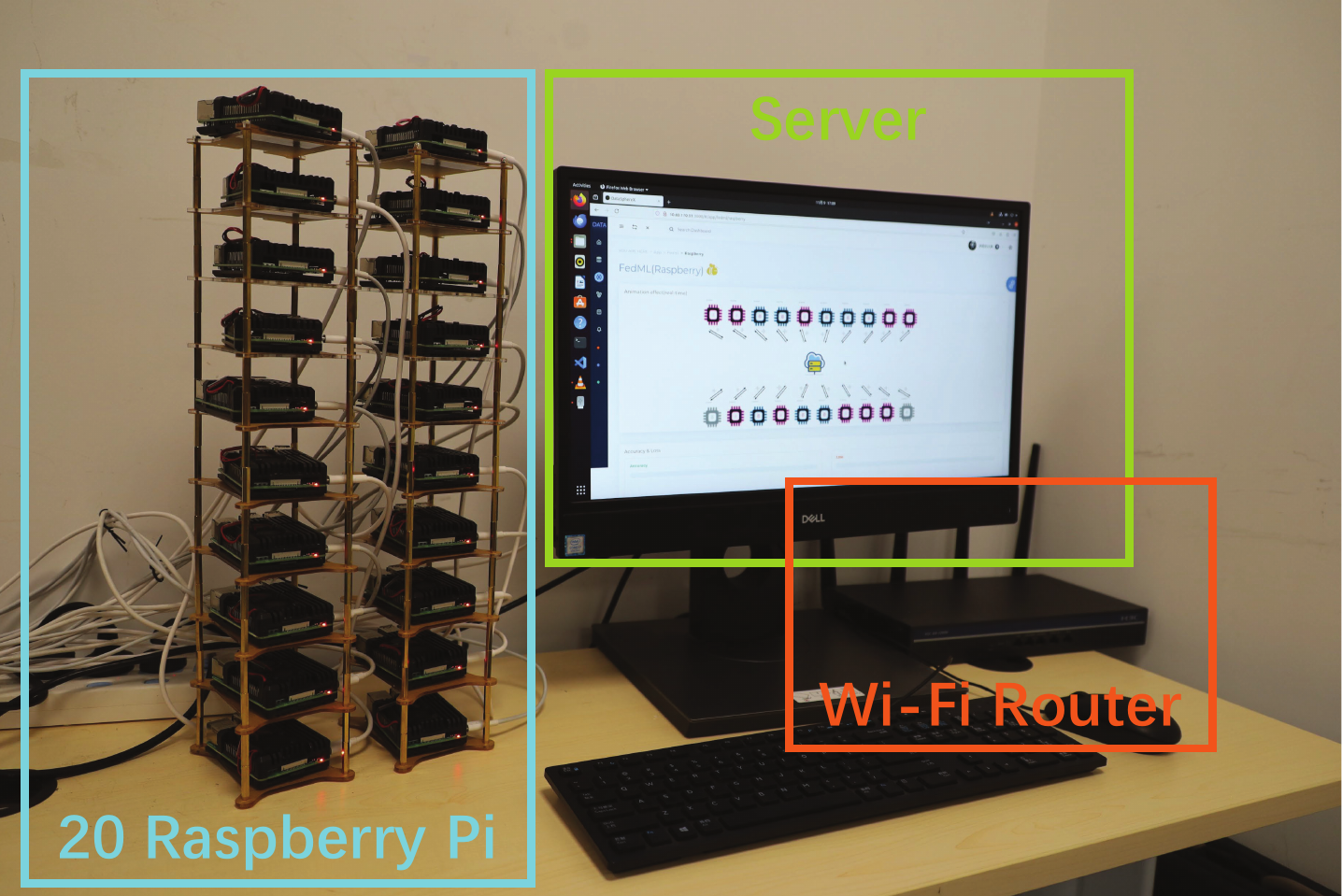}
     \vspace{-8pt}
    \caption{Hardware Prototype of FL.}
    \Description{Hardware Prototype of FL}
    \label{HD}
    \vspace{-12pt}
\end{figure}

To better characterize real-world applications, we carry out our experiments on a hardware prototype of FL, as shown in Figure \ref{HD}. The FL hardware prototype consists of $20$ Raspberry Pis that serve as clients, along with a desktop computer that functions as the central server for coordination (i.e., the platform in Figure \ref{framework}). The client and server communication occurs through a Wi-Fi router using a TCP-based socket interface.
\subsubsection{Heterogeneous Clients and Learning Implementation} We consider a set of $N=20$ heterogeneous clients interested in the FL model. For ease of illustration\footnote{While we only present experimental results under this setting due to the page limit, the insights for other function types of model utility and participation cost are similar.}, we define the FL model utility function of clients as $U(\epsilon)=40\cdot\epsilon^{-16}$, and the participation cost of clients is proportional to their training data size. To simulate the heterogeneity among clients, we categorize $20$ clients into $I=3$ types: $10$ type-$1$ clients, five types-$2$ clients, and five type-$3$ clients. Moreover, we use the widely adopted MNIST \cite{Lecun} and CIFAR-10 \cite{krizhevsky2009learning} data sets in our experiments\footnote{We also conduct experiments on the CIFAR-100 \cite{krizhevsky2009learning} and SVHN \cite{37648} data sets, which have similar results, as demonstrated in the online appendix \cite{Online_appendix}.}, which contain $70,000$ grayscale images of handwritten digits and $60,000$ color images of objects, respectively. Table \ref{CF10} summarizes the training data size $D_i$ of each client type $i\in\mathcal{I}$, as well as the testing data size used to evaluate the trained global FL model performance. While we focus on this small-scale setup, the insights are the same for a larger number of clients and more client types. Below, we will detail the implementation settings on both data sets.
\begin{table}[ht]\vspace{-8pt}
\caption{Training Data Size of Clients and Testing Data Size}\label{CF10}\vspace{-10pt}
    \centering
    \begin{tabular}{|c|c|c|c|c|}
    \hline
    \multirow{2}{*}{\textbf{Data Set}} & \multirow{2}{*}{\textbf{Testing Data Size}} & \multicolumn{3}{c|}{\textbf{Training Data Size}}\\ \cline{3-5} 
          &         & \multicolumn{1}{c|}{$D_1$}  & \multicolumn{1}{c|}{$D_2$}  & $D_3$  \\ \hline
    MNIST & $10000$ & \multicolumn{1}{c|}{$50$}   & \multicolumn{1}{c|}{$120$}  & $300$  \\ \hline
    CIFAR-10 & $10000$ & \multicolumn{1}{c|}{$1000$} & \multicolumn{1}{c|}{$2250$} & $5750$ \\ \hline
    \end{tabular}
    \vspace{-6pt}
\end{table}

\textbf{Setup with MNIST Data Set:} In the empirical experiments on the MNIST data set, we follow the same setup as in our theoretical analysis, and study extreme scenarios without data distribution heterogeneity, i.e., client variance $\sigma^2=0$. Specifically, we distribute data samples among $20$ clients in an i.i.d. fashion. Meanwhile, we adopt a convex multinomial logistic regression model, with initialized parameters $\boldsymbol{w}_0=\boldsymbol{0}$ and stochastic gradient descent (SGD) batch size $b=32$. Similar to \cite{Li2020On}, we use an initial learning rate $\eta_0=0.001$ with decay rate $\frac{\eta_0}{1+r}$, where $r$ is the communication round index, and the local iteration number $E=20$. The total number of communication rounds during the training process is $500$.

\textbf{Setup with CIFAR-10 Data Set:} In the empirical experiments on the CIFAR-$10$ data set, we generalize from our theoretical setting to explore more complex and realistic non-convex models. Meanwhile, we simulate a more practical FL scenario of non-i.i.d. data distribution. In particular, we assign one class of data samples to type-$1$ clients, five classes to type-$2$ clients, and all $10$ classes to type-$3$ clients. We employ a convolutional neural network (CNN) model, which consists of two convolutional layers (Relu activation function), three fully connected layers, and a max pooling layer applied to each convolutional layer. We initialize the CNN model with parameters $\boldsymbol{w}_0=\boldsymbol{0}$ and set the SGD batch size to $b=64$. The initial learning rate is $\eta_0=0.1$ with the decay rate $\frac{\eta_0}{1+r}$, and the local iteration number $E=30$. The total number of communication rounds during the training process is $600$.

For all experiments on both data sets, we evaluate the aggregated FL model in each communication round, where without loss of generality, we take the test loss as the generalization error $\epsilon$ of the trained FL model. Furthermore, to enhance reliability, our results are averaged over 30 experiments.

\subsection{Social Efficiency of the SEMTS Mechanism and MTS Framework}
Next, in Figure \ref{SBW} and Figure \ref{SW}, we evaluate the social efficiency of our proposed SEMTS mechanism and MTS framework, respectively. 
\subsubsection{Social Efficiency of SEMTS Mechanism} 
\begin{figure}[ht]
\vspace{-13pt}
    \centering
    \subfigure[Social Welfare (Linear Model and i.i.d. Data Distribution over MNIST)]{\label{M_BW}\includegraphics[width=0.236\textwidth]{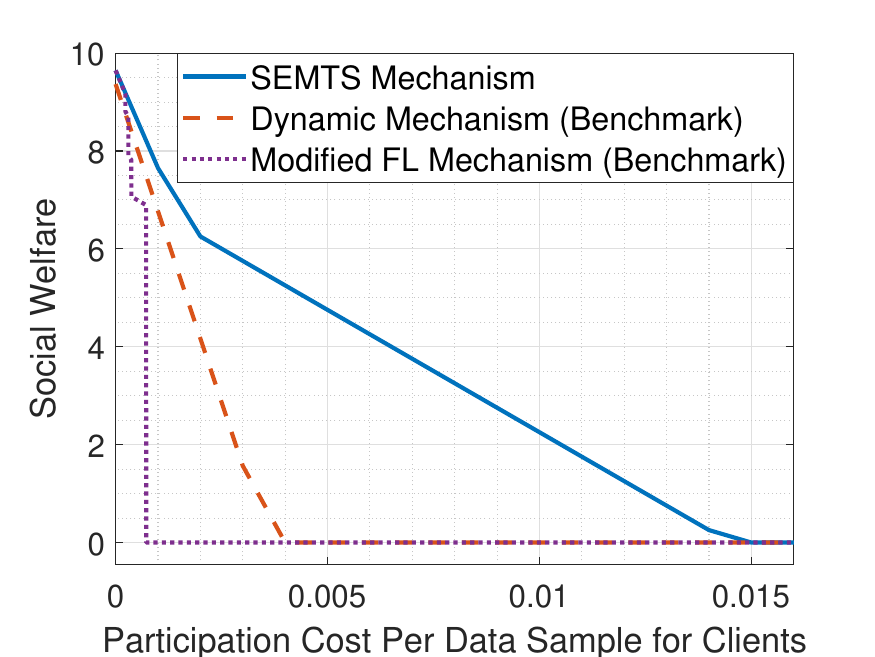}}
    \subfigure[Social Welfare (Non-convex Model and non-i.i.d. Data Distribution over CIFAR-10)]{\label{C_BW}\includegraphics[width=0.236\textwidth]{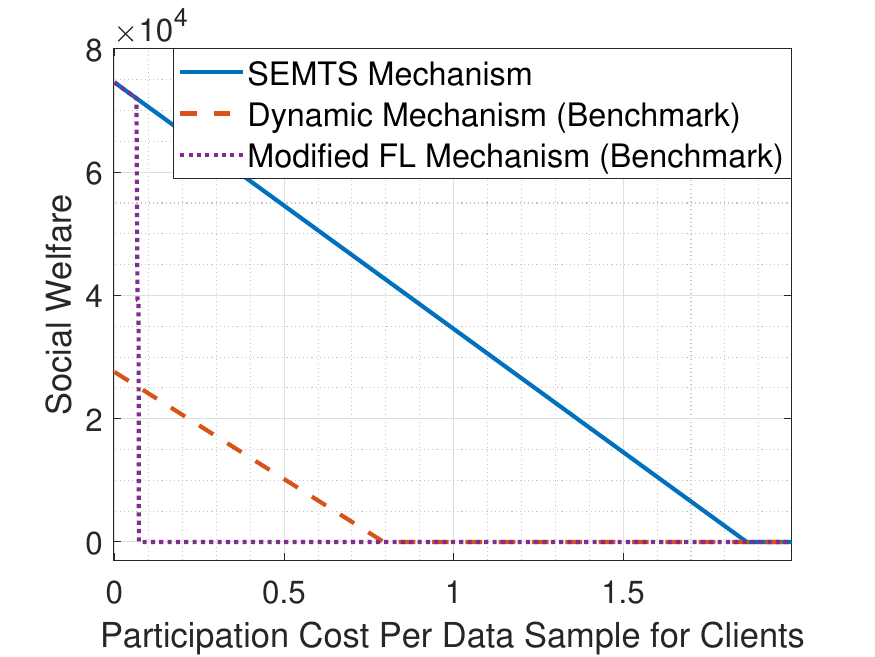}}
     \vspace{-10pt}
    \caption{Social Efficiency of Different Mechanisms under the MTS Framework.}
    \Description{Social Efficiency of Different Mechanisms under the MTS Framework}
    \vspace{-6pt}
    \label{SBW}
\end{figure}

In Figure \ref{SBW}, we compare the social welfare of our proposed SEMTS mechanism against that of two following benchmarks under the MTS framework:
\begin{itemize}
    \item Dynamic Mechanism \cite{hu2023}: The dynamic mechanism proposed in \cite{hu2023} considers clients' evolving expectations of FL model performance, i.e., a network effects game. However, incomplete information and client heterogeneity could lead this benchmark to deviate from social welfare maximization.
    \item Modified FL Mechanism \cite{Yan22}: This benchmark is a modified version of the existing socially efficient mechanism in \cite{Yan22}, in order to adapt to the MTS framework. Specifically, the modified FL mechanism sets the model price equal to the FL model utility but provides no participation reward, failing to deal with the network effects of client participation.
\end{itemize}

In the comparison of Figure \ref{SBW}, we illustrate how the social welfare of different mechanisms varies with the per-sample participation cost for all clients. From Figure \ref{M_BW} and \ref{C_BW}, as the participation cost increases, the social welfare of all mechanisms declines and eventually reaches zero. However, our proposed SEMTS mechanism achieves significantly higher social welfare than the benchmarks. In particular, benchmarks can only achieve a relatively competitive social welfare to our SEMTS mechanism at low participation costs. Once costs increase further, the benchmarks suffer a quick drop in social welfare due to the client heterogeneity and negative network effects of client participation. In contrast, the SEMTS mechanism can incentivize heterogeneous clients to overcome the temporary phase of negative network effects or negative payoff, maximizing social welfare by exploiting positive network effects. Consequently, our mechanism maintains high social welfare even with high participation costs of clients.

\subsubsection{Social Efficiency of MTS Framework}
\begin{figure}[ht]
\vspace{-13pt}
    \centering
    \subfigure[Social Welfare (Linear Model and i.i.d. Data Distribution over MNIST)]{\label{M_Wr}\includegraphics[width=0.236\textwidth]{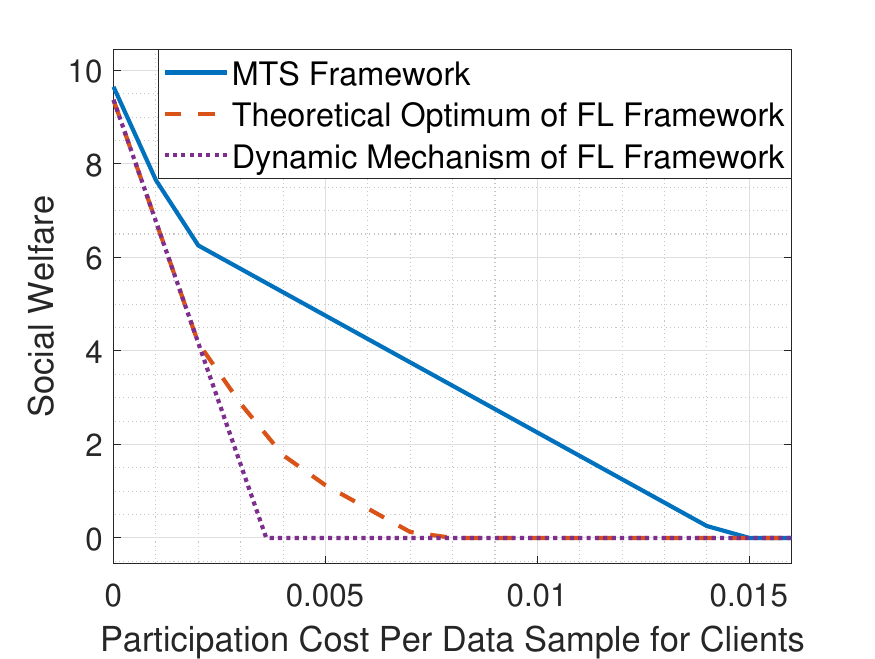}}
    \subfigure[Social Welfare (Non-convex Model and non-i.i.d. Data Distribution over CIFAR-10)]{\label{C_Wr}\includegraphics[width=0.236\textwidth]{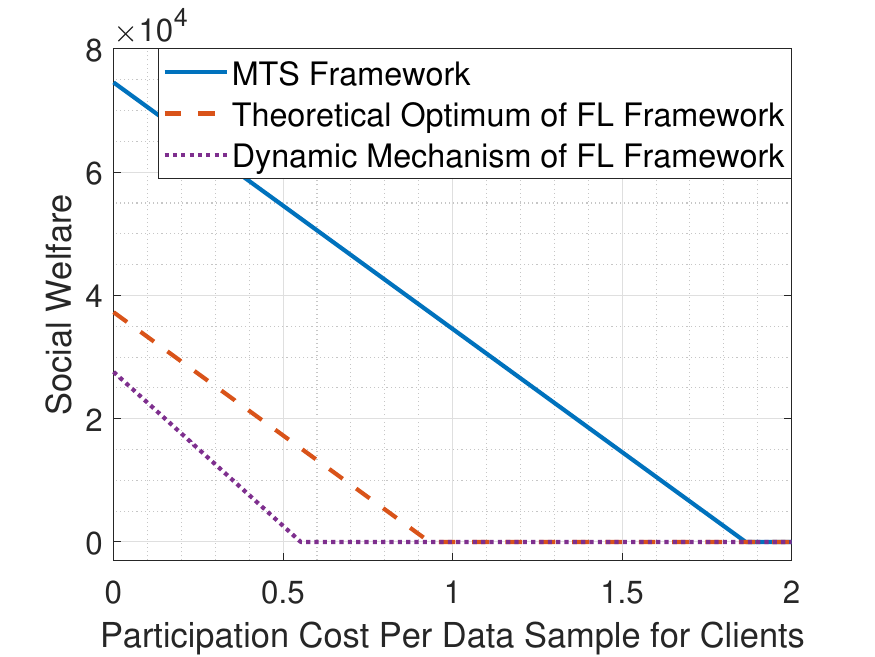}}
    \vspace{-10pt}
    \caption{Social Efficiency of Different Frameworks.}
    \Description{Social Efficiency of Different Frameworks}
     \vspace{-6pt}
    \label{SW}
\end{figure}

In Figure \ref{SW}, we further present a comparison between the social welfare achieved by our proposed mechanism under the MTS framework versus two benchmarks under the FL framework:
\begin{itemize}
    \item Theoretical Optimum of FL Framework: This benchmark implies the highest possible performance of the existing FL frameworks (e.g., \cite{mcmahan2017communication,Tan2022,Abdul2021}) and thus effectively serves as a theoretical upper bound of social welfare for comparison.
    \item Dynamic Mechanism \cite{hu2023} of FL Framework: Although the dynamic mechanism \cite{hu2023} under the FL framework can capture the dynamics of clients’ participation decisions to a certain degree, heterogeneous clients can only obtain and utilize the trained global FL model through participation, failing to maximize social welfare.
\end{itemize}

As shown in Figure \ref{SW}, with the rise of the per-sample participation cost for clients, the social welfare declines under both frameworks. However, the MTS framework introduces an extra purchasing option and leverages network effects through our designed SEMTS mechanism. This significantly increases the social welfare and makes it more resilient to rising participation costs. As a result, except when the social welfare is zero under both frameworks, the MTS framework achieves an overall social welfare that is $148.86\%$ and $300\%$ higher than that of the theoretical optimum under the FL framework on the MNIST and CIFAR-$10$ data set, respectively.

\section{Conclusion}
\label{S5}
This work provides the first study on social welfare optimization in FL with network effects. Our analysis reveals that the network effects of client participation are not monotonic, which depends on client heterogeneity. To improve social welfare, we propose the MTS framework for FL. By exploiting network effects, our designed SEMTS mechanism under the MTS framework can incentivize heterogeneous clients to maximize social welfare, relying only on payments from model customers without extra incentive costs.

In future works, we will extend our incentive mechanism to consider clients' strategic decisions of data contribution, such as limited participation using only part of their private data. Additionally, we will refine our theoretical model by exploring clients' utility and computational heterogeneity. Such advancements, while significantly complicating the equilibrium analysis and mechanism design, are expected to enhance the practicality of our research.
\begin{acks}
  This work is supported by National Natural Science Foundation of China (Project 62271434, 62102343), Guangdong Research (Project 2021QN02X778), Guangdong Basic and Applied Basic Research Foundation (Project 2021B1515120008, 2023A1515240051), Shenzhen Science and Technology Innovation Program (Project JCYJ202-10324120011032, JCYJ20230807114300001, JCYJ20220818103006012), Shenzhen Key Lab of Crowd Intelligence Empowered Low-Carbon Energy Network (No. ZDSYS20220606100601002), Shenzhen Stability Science Program 2023, and Shenzhen Institute of Artificial Intelligence and Robotics for Society.
\end{acks}

\bibliographystyle{ACM-Reference-Format}
\bibliography{ref}

\end{document}